# Scores of a specific field-normalized indicator calculated with different approaches of field-categorization: Are the scores different or similar?

Robin Haunschild*$, Angela D. Daniels**, and Lutz Bornmann*, ***

*$ Corresponding author*

* *R.Haunschild@fkf.mpg.de, L.Bornmann@fkf.mpg.de*

Max Planck Institute for Solid State Research, Heisenbergstr. 1, Stuttgart, 70569 (Germany)

** *ADaniels@cas.org*

CAS, a division of the American Chemical Society, Custom Services,

2540 Olentangy River Road, Columbus, Ohio 43202-1505 (USA)

*** *bornmann@gv.mpg.de*

Science Policy and Strategy Department, Administrative Headquarters of the Max Planck

Society, Hofgartenstr. 8, Munich, 80539 (Germany)

Abstract

Usage of field-normalized citation scores is a bibliometric standard. Different methods for field-normalization are in use, but also the choice of field-classification system determines the resulting field-normalized citation scores. Using Web of Science data, we calculated field-normalized citation scores using the same formula but different field-classification systems to answer the question if the resulting scores are different or similar. Six field-classification systems were used: three based on citation relations, one on semantic similarity scores (i.e., a topical relatedness measure), one on journal sets, and one on intellectual classifications. Systems based on journal sets and intellectual classifications agree on at least the moderate level. Two out of the three sets based on citation relations also agree on at least the moderate level. Larger differences were observed for the third data set based on citation relations and semantic similarity scores. The main policy implication is that normalized citation impact scores or rankings based on them should not be compared without deeper knowledge of the classification systems that were used to derive these values or rankings.

# Introduction[1]

It is one principle in the Leiden manifesto for the professional application of bibliometrics to use field-normalized scores instead of simple citation counts (Hicks, Wouters, Waltman, de Rijcke, & Rafols, 2015). These scores reflect the impact of papers against the backdrop of their reference sets – papers published at the same time and in the same field. An important topic in the calculation of these scores is the definition of fields, which are used as reference sets (Wilsdon et al., 2015; Wouters et al., 2015). Four different approaches of field-categorization are currently (mainly) used for normalizing impact without a clear preference for one alternative: (1) journal sets, (2) intellectual assignments, (3) citation relations, and (4) semantic similarity scores.

Waltman and van Eck (2019) reported Pearson correlations between field-normalized citation scores using Web of Science (WoS) subject categories and classifications based on citation relations for faculties, departments, and research groups at selected universities. They found correlations of 0.89 and higher as well as mean absolute score differences of 0.17 and lower. Scheidsteger, Haunschild, Hug, and Bornmann (2018) studied the concordance of field normalized scores for publications of a research institution focused on computer science using WoS and Microsoft Academic. Their normalization procedure used different classification schemes (WoS subject categories and fields of research from Microsoft Academic). They found a good agreement of the statistical concordance of the scores.

In this study, we compare normalized citation scores, which have been calculated based on the three approaches of field-categorization to build reference sets. We are interested whether they lead to the same, similar, or different scores for the same papers – if the formula for calculating the scores is held constant. Since all approaches are in current use for field-normalization in similar research evaluation contexts, we expect similar scores. Large differences between the scores could question the use of field-normalized scores in research evaluation, as long as no standard approach has been established. Such a study of possible differences is needed because the effect of the choice of a classification scheme on the values of the normalized indicators is largely unknown. When comparing normalized citation impact values of individual publications or aggregates thereof (e.g., scientists or universities) such effects should be known. If the effects are large, normalized values that were calculated using different classification schemes should not be compared with each other.

---

[1] This is a substantially extended version of Haunschild, Marx, French, and Bornmann (2018).

This study focuses on chemistry and related sciences utilizing a comprehensive dataset from CAS, a division of the American Chemical Society. CAS offers the largest database of the literature in these fields including intellectual assignments of fields to papers.

# Methods

## *Approaches of field-classification*

This study compares the agreement of normalized citation scores for the same papers, which have been calculated based on the following three field-categorization approaches:

(1) The most frequent approach in bibliometrics is to use subject categories that are defined by Clarivate Analytics for WoS or by Elsevier for Scopus to assign papers to fields. The subject categories pool journals to sets, which publish papers in similar research areas (e.g., biochemistry or economics). It is an advantage of journal sets that they define a multidisciplinary classification system covering all research areas (Wang & Waltman, 2016). It is a disadvantage of the sets that they stretch to their limits with multi-disciplinary journals (e.g., *Nature* or *Science*) and journals covering many subfields (e.g. *Physical Review Letters*, or *The Lancet*). These journals cannot be reliably and validly assigned to one field (Haddow & Noyons, 2013) on the journal basis. However, the papers of multidisciplinary journals can be reassigned on a paper-level (Evidence, 2010).

(2) To overcome the limitations of journal sets, Bornmann, Mutz, Neuhaus, and Daniel (2008) propose to use mono-disciplinary classification systems (Waltman, 2016), e.g., CAS sections in chemistry and related areas (Bornmann & Daniel, 2008; Bornmann, Schier, Marx, & Daniel, 2011), MeSH (Medical Subject Headings) terms in biomedicine (Bornmann, et al., 2008; Leydesdorff & Opthof, 2013; Strotmann & Zhao, 2010), or PACS (Physics and Astronomy Classification Scheme) codes in physics and related areas (Radicchi & Castellano, 2011). In these systems, experts in the field or the authors themselves assign each specific paper to the corresponding subfield, highlighting the most important aspects of the papers. It is an advantage of these systems that they have been introduced to reflect the subfield patterns in specific fields. Their disadvantage is that they can only be used for the normalization of papers from one discipline (and related areas).

(3) Waltman and van Eck (2012) introduced a multi-disciplinary classification system, which is based on direct citation relations between papers. The algorithm for computing the classification system needs three basic parameters as input in addition to the direct citation network: (i) the number of levels of the system, (ii) the resolution parameter, and (iii) the minimum number of papers per class (field). The approach is already in use in the Leiden ranking (see http://www.leidenranking.com/) for the calculation of normalized impact scores. The empirical results of Klavans and Boyack (2017) indicate that algorithmically constructed classifications are more accurate than classifications based on journal sets. Similar positive results have been published by Perianes-Rodriguez and Ruiz-Castillo (2016). Leydesdorff and Milojević (2015) criticize the classification system as follows: "Because these 'fields' are algorithmic artifacts, they cannot easily be named (as against numbered), and therefore cannot be validated. Furthermore, a paper has to be cited or contain references in order to be classified, since the approach is based on direct citation relations" (p. 201). It seems yet that Sjögårde, Ahlgren, and Waltman (2020) might have found a solution for this problem. Very recently,

Clarivate Analytics introduced an algorithmic classification with three levels based on WoS data (Szomszor, Adams, Pendlebury, & Rogers, 2021).

(4) Boyack and Klavans (2018) provided a publication classification system that is based on semantic similarity scores (i.e., a topical relatedness measure). The relatedness measures were based on words in titles and abstracts of PubMed publications along with their MeSH terms. They used PubMed records from 1975 onwards for the clustering procedure according to the topical relatedness measures because only very few records before 1975 contained abstracts.

*Statistics*

The overview of Waltman (2016) demonstrates that several different approaches of calculating field-normalized scores have been developed. In this study, we use the so-called normalized citation score (NCS) to compare field-normalized scores, since it is still the most frequently used approach. For the calculation of the NCS, each paper's citation count is divided by the average citation count in a corresponding reference set. The reference sets are defined by the papers, which belong to the same field (as defined by the field categorization approach) and publication year as the focal paper. If, for example, the paper has 3 citations and the average in the field is 10.67, the NCS of the paper is 3/10.67=0.28. The NCS is formally defined as

$$NCS = \frac{c_i}{e_i}$$

where $c_i$ is the citation count of a focal paper and $e_i$ is the corresponding citation rate in the field (Lundberg, 2007; Rehn, Kronman, & Wadskog, 2007; Waltman, van Eck, van Leeuwen, Visser, & van Raan, 2011). Since the number of citations received by a paper depends on the time since publication, the NCS is calculated for publications from the same year. Using the different approaches of field-categorization, we calculated six NCS for every paper: $NCS_{WoS}$ (based on WoS journal sets), $NCS_{CAS}$ (based on CAS sections), $NCS_{L15}$, $NCS_{L12\_2lvl}$, $NCS_{L12\_3lvl}$ (all three based on citation relations), and $NCS_{ST}$ (based on topical relatedness measures). Statistical analysis was done using R (R Core Team, 2018).

In this study, we are interested in the relationship between $NCS_{WoS}$, $NCS_{CAS}$, $NCS_{L15}$, $NCS_{L12\_2lvl}$, $NCS_{L12\_3lvl}$, and $NCS_{ST}$ to investigate the extent of agreement and disagreement between the different NCS values. We group the papers in our dataset according to the Characteristics Scores and Scales (CSS) method. This method was proposed by Glänzel, Debackere, and Thijs (2016) for normalizing citation counts. However, we do not use the CSS method here for normalizing the citation counts, only for grouping the papers according to their citedness as represented by their NCS value. For each NCS separately ($NCS_{WoS}$, $NCS_{CAS}$, $NCS_{L15}$, $NCS_{L12\_2lvl}$, $NCS_{L12\_3lvl}$, and $NCS_{ST}$), the CSS classifications are obtained by (1) truncating the publications at their mean NCS and (2) recalculating the mean of the truncated part. Performing this procedure three times leads to four impact classes. Following Glänzel, et al. (2016), we labeled the four classes with "poorly cited", "fairly cited", "remarkably cited", and "outstandingly cited". The poorly cited papers are below the average impact of all papers; the other three classes are above this average and further differentiate the papers in the high impact sectors.

We undertook 15 pairwise comparisons to investigate the differences between the three NCS variants. Each pair is compared in a 4 x 4 contingency table. The cells in the diagonal of the table reveal the papers, which have been assigned to a CSS class in agreement of both NCS, and the share of papers assigned in agreement are calculated. This procedure leads to a similarity measure referred to as "level of agreement". We further calculated the weighted Kappa coefficient in this study, which is a robust alternative to the share of agreement, since the possibility of agreement occurring by chance is taken into consideration (Gwet, 2014). Table 1 shows the weights that we used in the calculation of the Kappa coefficient. The weighted Kappa considers that disagreements between the various categories are different. For example, the difference between "Poorly cited" and "Outstandingly cited" is larger (and should have less weight) than the difference between "Poorly cited" and "Fairly cited" (which should have more weight). The weights in Table 1 are frequently used in studies measuring agreement with weights. The weighted Kappa coefficients were calculated using the R package 'irr' (Gamer, Lemon, & Singh, 2019).

A further advantage of using the Kappa coefficient is that guidelines by Landis and Koch (1977) are available for the proper interpretation of the level of agreement: <0.00 "poor", 0.00-0.20 "slight", 0.21-0.40 "fair", 0.41-0.60 "moderate", 0.61-0.80 "substantial", and 0.81-1.00 "almost perfect".

Table 1. Weights for the Kappa coefficient

|  | Poorly cited | Fairly cited | Remarkably cited | Outstandingly cited |
|---|---|---|---|---|
| Poorly cited | 1 | 0.75 | 0.5 | 0.25 |
| Fairly cited | 0.75 | 1 | 0.75 | 0.5 |
| Remarkably cited | 0.5 | 0.75 | 1 | 0.75 |
| Outstandingly cited | 0.25 | 0.5 | 0.75 | 1 |

We additionally calculated concordance coefficients for continuous variables following Lin (1989, 2000) to measure the agreement between $NCS_{WoS}$, $NCS_{CAS}$, $NCS_{L15}$, $NCS_{L12\_2lvl}$, $NCS_{L12\_3lvl}$, and $NCS_{ST}$. Lin's concordance coefficient for sets x and y of n values each is calculated as follows:

$$lcc = \frac{{}^2/_n \sum_i (y_i - \bar{y})(x_i - \bar{x})}{{}^1/_n \sum_i (x_i - \bar{x})^2 + {}^1/_n \sum_i (y_i - \bar{y})^2 + (\bar{x} - \bar{y})^2}$$

where $\bar{x}$ and $\bar{y}$ are the average values of the sets x and y. For example, suppose we have the two sets x = (1, 2, 3, 4, 5, 6, 7, 8, 9, 10) and y = (11, 12, 13, 14, 15, 16, 17, 18, 19, 20). We obtain a perfect correlation but a concordance coefficient according to Lin of 0.142. Lin's concordance coefficients were calculated using the R packages 'DescTools' (Signorell et al., 2019).

We abstained from calculating correlation coefficients in this study, because we are interested in the agreement between two NCS (Lowenstein, Koziol-McLain, & Badgett, 1993). Correlation is a poor substitute for agreement. For example, systematic bias might be ignored. Suppose $NCS_{CAS}$, and $NCS_{WoS}$ have a perfect correlation (see the example of x and y above), but the $NCS_{CAS}$ consistently measures citation impact 0.5 levels lower than the $NCS_{WoS}$. In this case, the NCS would measure citation impact very differently despite the high correlation between both NCS variants.

*Data sets used*

Database for calculating $NCS_{WoS}$: The WoS journal sets are available in our in-house database developed and maintained by the Max Planck Digital Library (MPDL, Munich) and derived from the Science Citation Index Expanded (SCI-E), Social Sciences Citation Index (SSCI), Arts and Humanities Citation Index (AHCI) provided by Clarivate Analytics (Philadelphia, Pennsylvania, USA). These journal sets are grouped into 255 WoS subject categories. We calculated the $NCS_{WoS}$ values by using the journal sets and the citation counts from the WoS in-house database. The journal set classification of the WoS, however, assigns multiple fields to many publications without any priority. Therefore, we calculated for every paper an average of the $NCS_{WoS}$ values in each field to receive an overall score.

Database for calculating $NCS_{CAS}$: The $CAplus^{SM}$ database is an integrated source of journal articles and patent documents in many scientific disciplines. For the purpose of this study, over 12,000,000 journal publications published between 2000 and 2014 were used. CAS uses a hierarchical field classification scheme to assign the publications into five broad headings of chemical research (section headings), which are further separated into 80 scientific subject areas named as CAS Sections plus a separate CAS Section for unclassified documents. Most publications are assigned to only one section based on the main subject field; some publications are also assigned to a secondary section. To avoid multiple classifications of publications in this study, only the primary section assignment is used following previous studies (Bornmann & Daniel, 2008; Bornmann, et al., 2011). The section assignments are human-curated by scientists at CAS with specialized knowledge that lets them accurately extract and verify data and insights from each publication. The classification does not seem to be affected by the "indexer effect": according to Braam and Bruil (1992), the indexer classification accords with author preferences for 80% of the publications. We calculated the $NCS_{CAS}$ values using the CAS sections and the citation counts from the CAplus database.

Database for calculating the $NCS_L$ variants ($NCS_{L15}$, $NCS_{L12\_2lvl}$, and $NCS_{L12\_3lvl}$): The algorithmically constructed classifications and the algorithm itself by Waltman and van Eck (2012) have been made freely available. The algorithm can be run with and without hierarchical constraint. The classifications are uniquely assigned to papers: Each paper is assigned to only one classification. The hierarchical classifications are available on three different levels. We used the second (L12_2lvl) and third level (L12_3lvl). We refer to the non-hierarchical variant as L15. We downloaded the classifications of the papers and the corresponding WoS UTs for L12_2lvl and L12_3lvl on November 7th, 2014 from http://www.ludowaltman.nl/classification_system and for L15 on 28 May 2015 from https://www.leidenranking.com/information/fields. These three classification systems have a different number of clusters: (i) L12_2lvl has 672 clusters, (ii) L15 has 3822 clusters, and (iii) L12-3lvl has 22412 clusters. The classifications were matched via the WoS UT to the data in our in-house database. The $NCS_{L15}$, $NCS_{L12\_2lvl}$, and $NCS_{L12\_3lvl}$ values have been calculated by using these classifications and the citation counts from our WoS in-house database. The L12 classification system covers WoS publications from 2001 until 2011 while the L15 system covers WoS publications between 2006 and 2013.

Database for calculating $NCS_{ST}$: Boyack and Klavans (2018) freely provided a publication classification system based on topical relatedness measures. The relatedness measures were based on words in titles and abstracts of PubMed publications along with their MeSH terms. We downloaded the classification system on 22 October 2019 from https://www.scitech-strategies.com/pubmed-model-created-using-nih-sbir-funding. In this classification system, PubMedIDs (n=16,261,085) are assigned to 35,966 different clusters. Of those PubMedIDs, 6,565,130 belong to papers published between 2006 and 2011 (overlap of the algorithmically constructed classification systems L12 and L15, see above). The classifications were matched via the PubMedID to the data in our in-house database. The $NCS_{ST}$ values have been calculated by using this classification system and the citation counts from our WoS in-house database.

Matching of the various NCS values: The $NCS_{CAS}$ values for each paper were matched with the $NCS_{WoS}$ values via the DOI. The $NCS_{ST}$ values were matched with the $NCS_{WoS}$ values via the PubMedID. The other NCS values were matched with the $NCS_{WoS}$ values via the WoS UT. Selected analyses have been performed on single matches (e.g., match of $NCS_{WoS}$ with $NCS_{L15}$), too, for avoiding a bias in our results by reduction of our dataset due to the full matching procedure of all studied classification systems. We found only minor differences in the results of the single and full matches of the analyzed datasets. Matched publications with unique DOI and PubMedID (n=256,743) have been used in our main results. Despite the focus of CAplus on chemistry and ST being constructed within PubMed, most classifications are within our data set: 159 WoS subject categories (62.4%), 670 of the clusters in L12_2lvl (99.7%), 14683 of the clusters in L12_3lvl (65.5%), and 3098 (81.1%) of the clusters in L15 remain after matching with the two field-specific classification systems.

The WoS subject categories and the corresponding number of papers that are in our matched data set with at least 5000 papers are shown in Table 2. Besides chemical WoS subject categories, many biological and a few physical WoS subject categories can be observed in this list of most frequently occurring classifications in our data set.

Table 2: WoS subject categories (28 out of 255) with at least 5000 papers within our matched data set

| WoS subject category | Number of papers |
| --- | --- |
| Biochemistry & Molecular Biology | 38535 |
| Chemistry, Multidisciplinary | 21016 |
| Pharmacology & Pharmacy | 17433 |
| Chemistry, Physical | 17400 |
| Cell Biology | 16315 |
| Biotechnology & Applied Microbiology | 14330 |
| Oncology | 13610 |
| Neurosciences | 12287 |
| Biochemical Research Methods | 12225 |
| Environmental Sciences | 11190 |
| Genetics & Heredity | 11027 |
| Chemistry, Analytical | 10490 |
| Microbiology | 10431 |
| Immunology | 9874 |
| Chemistry, Organic | 9216 |
| Endocrinology & Metabolism | 9054 |
| Biophysics | 8594 |
| Materials Science, Multidisciplinary | 7843 |
| Hematology | 7606 |
| Chemistry, Medicinal | 7271 |
| Physics, Atomic, Molecular & Chemical | 7098 |
| Plant Sciences | 6742 |
| Toxicology | 6636 |
| Multidisciplinary Sciences | 6360 |
| Nanoscience & Nanotechnology | 6151 |
| Medicine, Research & Experimental | 5994 |
| Food Science & Technology | 5754 |
| Physiology | 5443 |

Figure 1 shows the distribution of the number of papers within our matched data set across the classifications of the six different classification systems visualized as a boxplot. The black bars show the median number of papers per classification. The black dots mark the outliers. One can see from the figure that WoS, CAS, and L12_2lvl are rather similar in paper density, although L12_2lvl has a slightly smaller paper density than WoS and CAS. WoS has a larger variation of paper densities across the fields than CAS which can be expected when the overlap of a multi-disciplinary classification system with a mono-disciplinary classification system is analyzed. Also, L12_3lvl, L15, and ST form a group of rather similar paper densities.

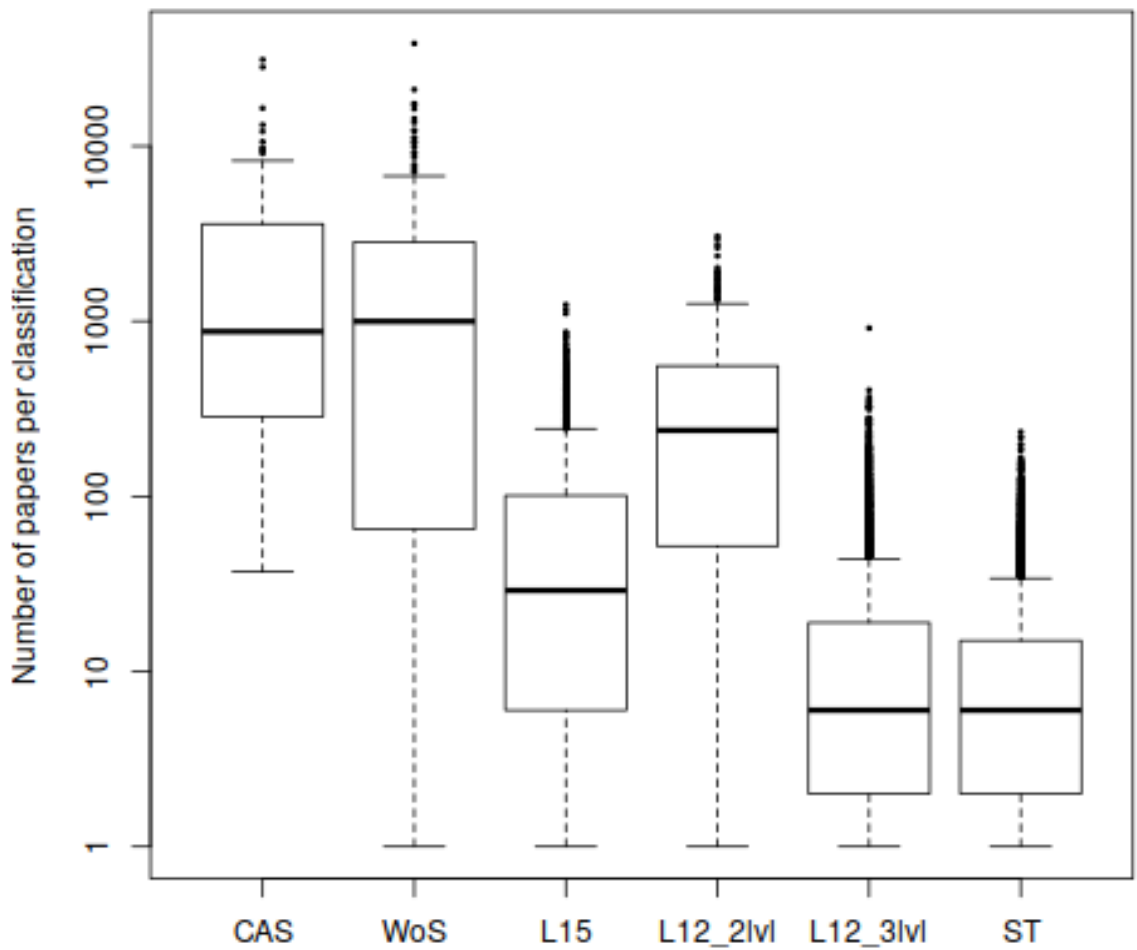

Figure 1: Boxplot of the number of papers across the classification systems within the matched data set.

## Results

Figure 2 visualizes the change of CSS assignment of the analyzed papers as an alluvial diagram. The alluvial diagram was produced using the R package 'alluvial' (Bojanowski & Edwards, 2016). The most changes can be observed between CSS classes 1 and 2. The color is assigned to the CSS classes: 1 in gray, 2 in red, 3 in green, and 4 in blue. Papers that stay in the same CSS class are represented by a band of the color of the class. If papers change the CSS class from one classification system to another, they are represented by a multi-color band (e.g., gray and red if the papers change from CSS class 1 to CSS class 2). The broader the band, the more papers are represented. In the following analysis, we observed that there was a reduction in similarity measures due to many changes of CSS classes.

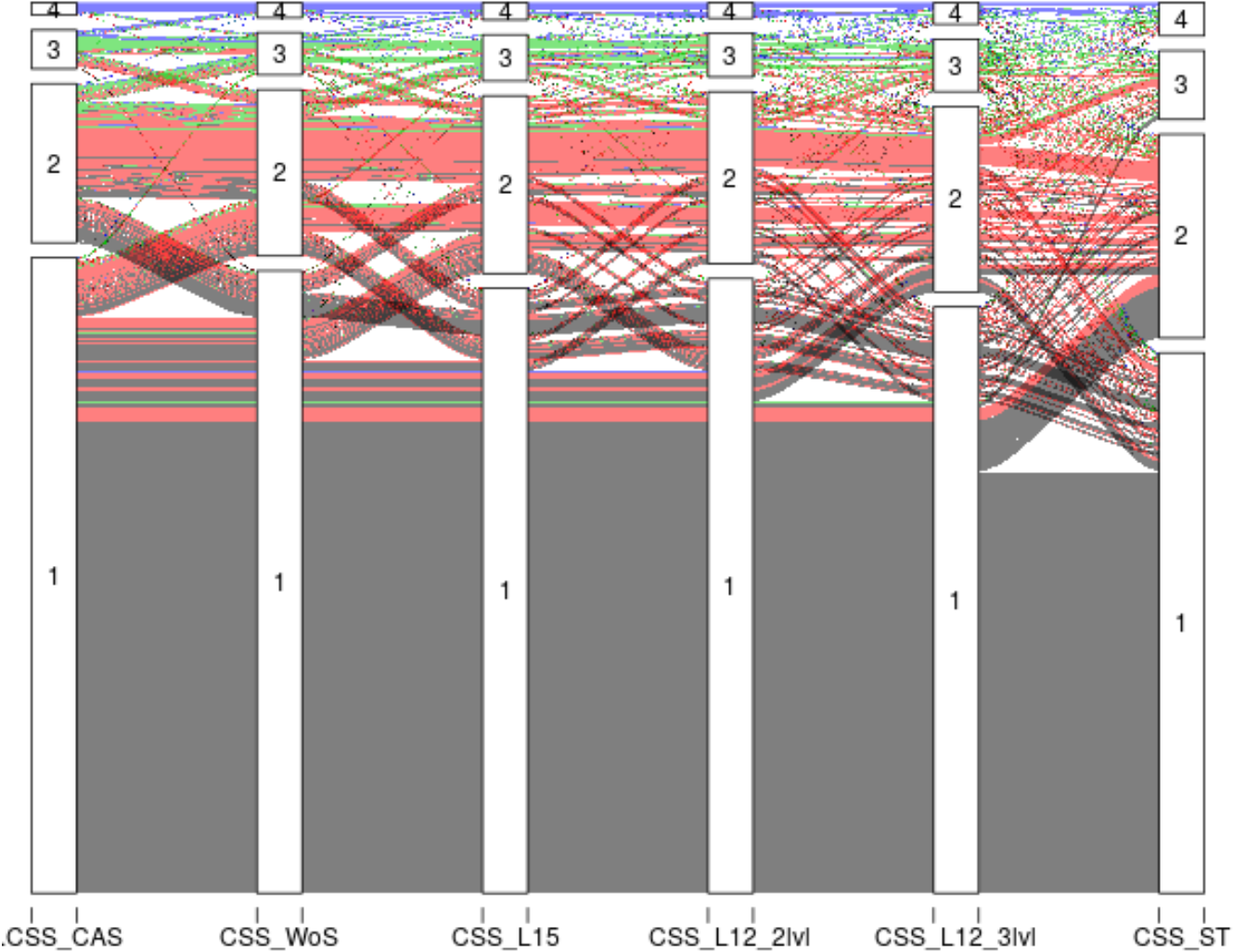

Figure 2: Alluvial diagram for the six different classification

Figure 3 shows the levels of agreement between the NCS scores of the six classification systems. The level of agreement is measured by the percentage of papers within the same class. The highest levels of agreement can be observed between L12, L15, WoS, and CAS: L15-

L12_2lvl > WoS-L12_2lvl > L15-L12_3lvl > CAS-WoS > L12_2lvl-L12_3lvl > CAS-L12_2lvl > WoS-L15 > CAS-L15 > WoS-12_3lvl > CAS-L12_3lvl. The SciTech classification system shows a lower level of agreement with the other classification systems than the other classification systems among each other: L15-ST > L12_3lvl-ST > L12_2lvl-ST > WoS-ST > CAS-ST.

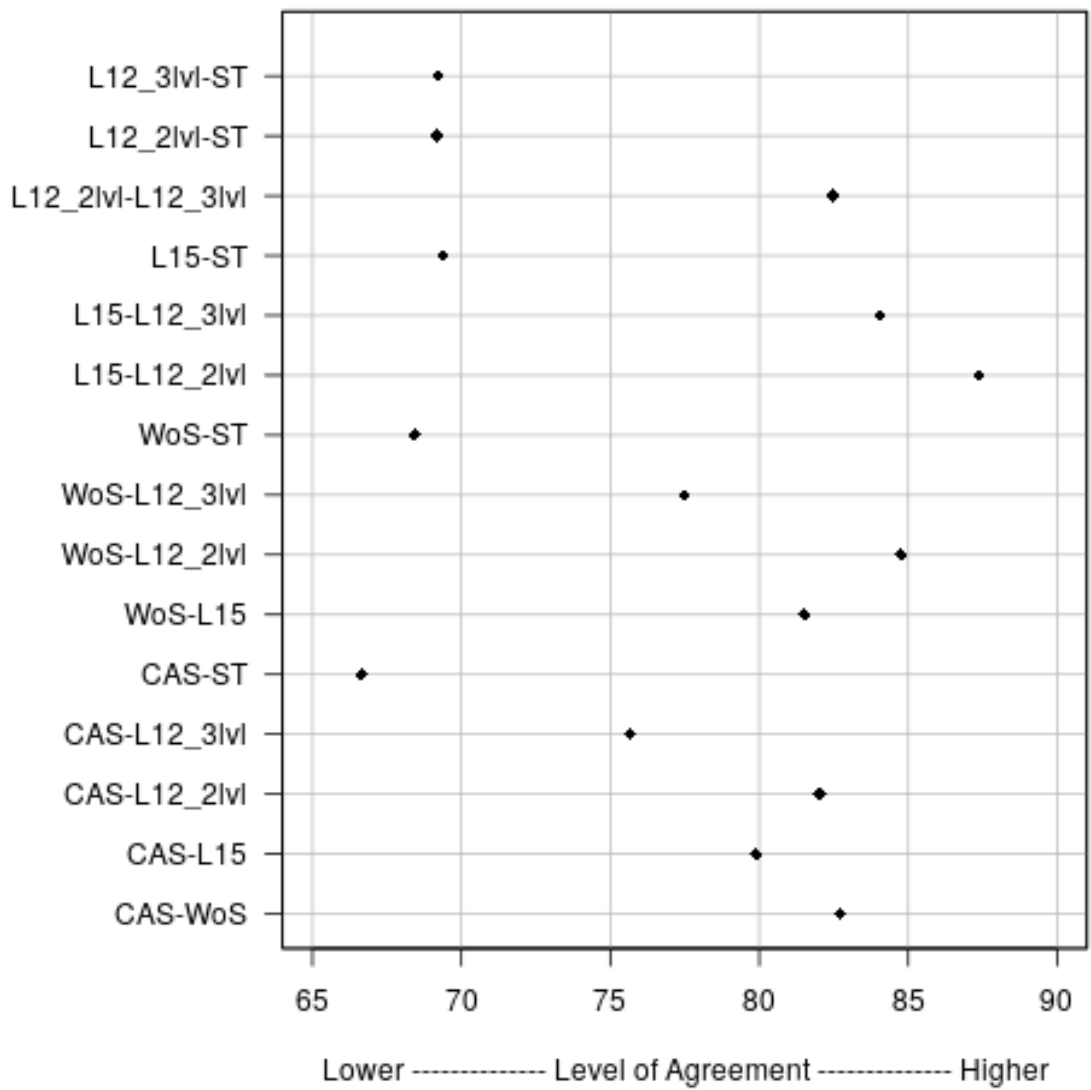

Figure 3: Levels of agreement between the six different classification systems that lead to 15 different combinations

Our results do not seem to be affected by the data reduction due to the multiple matches with the partly overlapping publication sets of the multiple classification systems. We calculated the level of agreement for selected combinations with dual matches only, for example: The multiple match shows a level of agreement of 81.53% for the pair WoS-L15 based on 256,743 publications, and the single match (WoS subject categories with L15 clusters) shows a level of agreement of 81.36% for the same pair based on 7,953,992 publications. Such comparisons indicate that our results are not biased towards the fields covered by the mono-disciplinary databases CAplus and PubMed.

Figure 4 shows the weighted Kappa coefficients for the combinations of the six different classification systems. The large circle is centered on the point estimate, and the lines inside the circles indicate the size of the 95% confidence interval. The circles and the lines are color-coded according to the interpretation guidelines by Landis and Koch (1977): Red indicates a low level of agreement, yellow a moderate level, and green a substantial level of agreement. The weighted Kappa coefficients show similar results like the levels of agreement. The interpretation guidelines by Landis and Koch (1977) provide the following grouping: (i) substantial agreement of L15-L12_2lvl > L15-L12_3lvl > WoS-L12_2lvl > L12_2lvl-L12_3lvl > WoS-L15 > CAS-WoS > CAS-L12_2lvl > CAS-L15, (ii) moderate agreement of WoS-12_3lvl > CAS-L12_3lvl > L12_3lvl-ST > L15-ST > L12_2lvl-ST > WoS-ST, and (iii) low agreement of CAS-ST. Only the 95% confidence intervals of WoS-L15 and CAS-WoS slightly overlap.

Our results do not seem to be affected by the data reduction due to the multiple matches with the partly overlapping publication sets of the different classification systems. We calculated the Kappa coefficients for selected combinations with single matches only, for example: The multiple match shows a Kappa coefficient of 0.66 for the pair WoS-L15 based on 256,743 publications, and the single match (WoS subject categories with L15 clusters) shows a Kappa coefficient of 0.68 for the same pair based on 7,953,992 publications. Such comparisons indicate that our results are not biased towards the fields covered by the mono-disciplinary databases CAplus and PubMed.

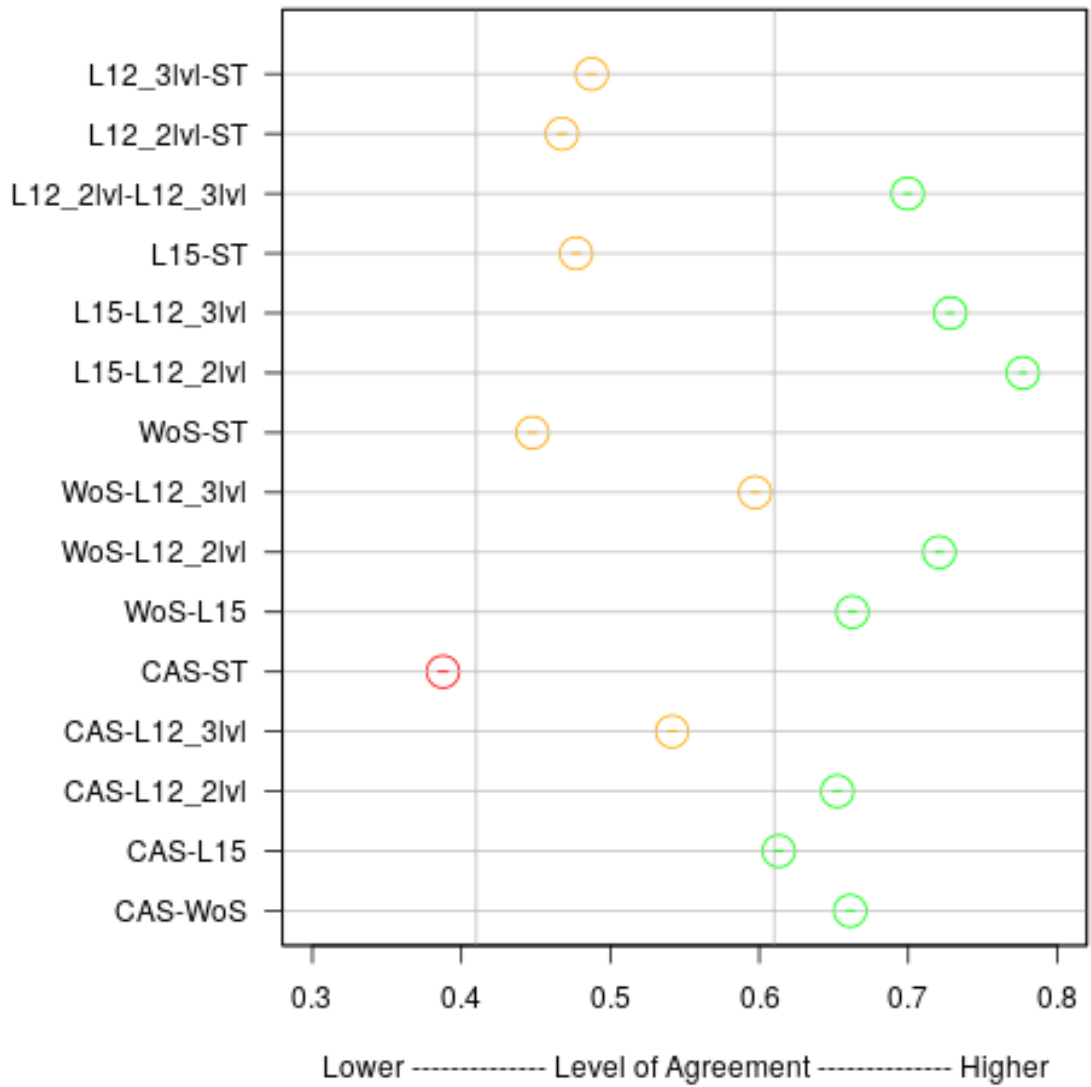

Figure 4: Weighted Kappa coefficients for the combinations of the six different classification systems that lead to 15 different combinations

Figure 5 shows Lin's concordance coefficient for the combinations of the six different classification systems. The large circle is centered on the point estimate, and the lines inside the circles indicate the size of the 95% confidence interval. Although the same broad interpretation can be reached for Lin's concordance coefficient as for the levels of agreement and kappa coefficients, the order is slightly different: L15-L12_2lvl > L15-L12_3lvl > CAS-WoS > L12_2lvl-L12_3lvl > WoS-L12_2lvl > WoS-L15 > CAS-L12_2lvl > CAS-L15 > WoS-12_3lvl > L12_3lvl-ST > CAS-L12_3lvl > L15-ST > L12_2lvl-ST > WoS-ST > CAS-ST. None of the 95% confidence intervals are overlapping.

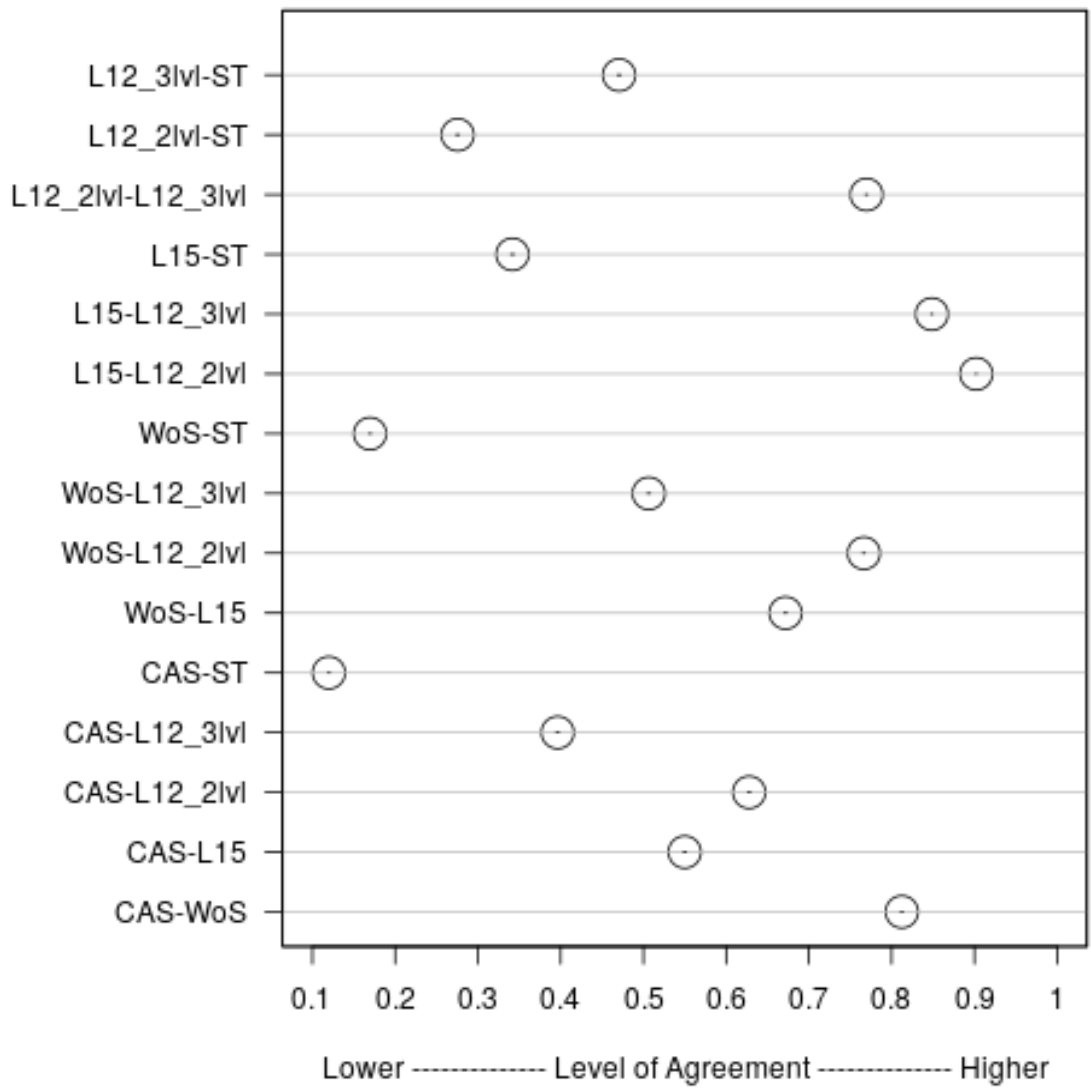

Figure 5: Lin's concordance coefficients for the combinations of the six different classification systems that lead to 15 different combinations

Our results do not seem to be affected by the data reduction due to the multiple matches with the partly overlapping publication sets of the different classification systems. We calculated Lin's concordance coefficients for selected combinations with single matches only, for example: The multiple match shows a concordance coefficient of 0.67 for the pair WoS-L15 based on 256,743 publications, and the single match (WoS subject categories and L15 clusters) shows a concordance coefficient of 0.65 for the same pair based on 7,953,992 publications. Such comparisons indicate that the agreement of WoS and L15 is slightly higher in the fields covered by the mono-disciplinary databases CAplus and PubMed than in the single match of the multidisciplinary classification systems.

## Discussion

According to Ioannidis, Boyack, and Wouters (2016) "the basic premise of normalization is that not all citations are equal. Therefore, normalization can be seen as a process of benchmarking". Although it is standard in bibliometrics to use field-normalized citation scores for cross-field comparisons (of universities, for example), different approaches exist for calculating these scores. The differences refer either to the method of calculating the scores (percentiles have been proposed as an alternative to scores based on average citations, Bornmann & Marx, 2015) or to the approach of field categorization which are used to build the reference set for each paper. In this study, we addressed the second aspect by comparing the normalized scores, which have been calculated based on three different approaches.

The analysis of the scores basically reveals an agreement which is at least at the moderate level except for the comparison of SciTech with CAS. Since we used the same method for calculating the scores based on the different approaches, the moderate level is lower than the level that we expected. The parallel use of the different approaches in the current research evaluation practice should have led to a generally higher level of agreement. The main policy implication is that normalized citation impact scores or rankings based on them should not be compared without deeper knowledge of the classification systems that were used to derive these values or rankings. However, our results also show that normalized scores based on intellectual field assignments are more in agreement with scores based on journal sets than with scores based on citation relations or semantic similarity scores. Thus, one can expect more similar scores based on intellectual assignments and journal sets than on algorithmically constructed classification systems. The reason for the similarity might be that intellectual assignments and journals are better rooted in the disciplines than virtual constructs based on algorithmically constructed classification systems. Another possible explanation is that classification systems of similar granularity are more likely to produce similar results. The WoS journal sets and the intellectual assignments by CAS have overall a similar granularity and show a substantial agreement. However, the classification systems L12_3lvl, and ST show only a moderate agreement although they have a similar granularity. CAS Sections were developed by CAS scientists with specialized knowledge in scientific disciplines and used to accurately index data and insights from each publication. According to Sugimoto and Weingart (2015), the establishment of new journals is a sign of emerging new disciplines. This might help to explain the similar results for WoS journal sets and CAS sections.

The results of this study should be interpreted against the backdrop that the main results of this study focus on one discipline only: chemistry and related areas. However, selected comparisons without the focus on chemistry and related areas showed very similar results. Furthermore, other statistical analyses could be performed. It is not clear whether our results can be generalized. Thus, we encourage similar studies with data from other disciplines using different statistical methods and as many classification schemes as possible.

**Acknowledgments**

Pennsylvania, USA). We would like to thank the Centre for Science and Technology Studies (CWTS) and SciTech Strategies for making their assignments to Web of Science (WoS) UTs or PubMedIDs available. Parts of this work using an older data set were performed during a research visit of one of the co-authors (RH) with CAS in Columbus, OH, USA). RH thanks CAS for support during his stay.